\documentclass[12pt]{article}
\usepackage{latexsym}
\usepackage{amsmath}
\usepackage{amssymb}
\usepackage{mathrsfs}
\usepackage{theorem}

\usepackage{cite}
\newtheorem{theorem}{Proposition}
{\theorembodyfont{\rmfamily}
\newtheorem{remark}{Remark}}
\date{}
\begin{document}
 \begin{center}
{\Large \textbf{General heavenly equation governs anti-self-dual gravity}}

\vspace{5mm}{\large\textbf{A. A. Malykh$^1$ and M. B.
Sheftel$^{2,3}$}}
\end{center}
 \vspace{2mm} $^1$ Department of Numerical Modelling, Russian State
Hydrometeorlogical  \\ $\phantom{^1 }$  University, Malookhtinsky
pr. 98, 195196 St. Petersburg, Russia \\
 $^2$ Department of Physics, Bo\u{g}azi\c{c}i
University 34342 Bebek, Istanbul, Turkey  \\
 $^3$ Department of Mathematics, North-West State Technical University\\
$\phantom{^1 }$ Millionnaya St. 5, 191186, St.
Petersburg, Russia \\
 \vspace{1mm}
 \\ $\phantom{^1 }$ E-mail: andrei-malykh@mail.ru,
                            mikhail.sheftel@boun.edu.tr
\begin{abstract}\noindent
 We show that the general heavenly equation, suggested recently by Doubrov and Ferapontov \cite{fer},
governs anti-self-dual (ASD) gravity. We derive ASD Ricci-flat vacuum metric governed
by the general heavenly equation, null tetrad and basis of 1-forms for this metric.
We present algebraic exact solutions of the general heavenly equation as a set of zeros of
homogeneous polynomials in independent and dependent variables. A real solution is obtained for
the case of neutral signature.
\end{abstract}

\section{Introduction}
\setcounter{equation}{0}
 \label{sec-intro}

There are several four-dimensional scalar equations of Monge-Amp\`ere type
that determine potentials of ASD Ricci-flat metrics: first and second heavenly equations of Pleba\~nski \cite{pleb},
complex Monge-Amp\`ere equation ($CMA$) as a real version of the first heavenly equation and Husain equation \cite{husain}
together with the closely related mixed heavenly equation \cite{shma}. All these equations appear as canonical forms
of the general second-order four-dimensional equation which admits partner symmetries
and turns out to be of Monge-Amp\`ere type with the additional constraints
that it has a two-dimensional divergence form (instead of four-dimensional divergence form which it will have in general)
and contains only second partial derivatives of the unknown \cite{shma}. In a recent paper \cite{fer}, Doubrov and Ferapontov
classified all integrable four-dimensional Monge-Amp\`ere equations with no restriction of admitting a two-dimensional divergence form.
Among resulting normal forms of these equations they presented only one equation that is not of a two-dimensional divergence form
(it has a three-dimensional divergence form), which they called \textit{general heavenly equation}
\begin{equation}\label{gener}
  \alpha u_{12}u_{34} + \beta u_{13}u_{24} + \gamma u_{14}u_{23} = 0,\qquad \alpha + \beta + \gamma = 0,
\end{equation}
where $\alpha, \beta$ and $\gamma$ are arbitrary constants with one linear dependence between them. Here and further on, subscripts mean partial derivatives
with respect to variables $z^1,z^2,z^3,z^4$. All other normal forms are either well-known equations
that govern self-dual gravity or the ``modified heavenly equation'' which is a particular case of our
``asymmetric heavenly equation'' \cite{shma}. Equation (\ref{gener}) also appeared in our recent paper \cite{mash}, where it was shown that by using partner symmetries one can obtain its particular solutions, which also satisfy complex Monge-Amp\`ere equation.

Our original motivation for this study of the general heavenly equation was just to make sure that all four-dimensional equations
of the Monge-Amp\`ere type can be used to describe anti-self-dual gravity. Then we realized an importance of the characteristic feature of the general heavenly equation that being homogeneous, in contrast to other heavenly equations mentioned above, it enables one to construct easily algebraic solutions, sets of zeros of polynomials in dependent and independent variables. By ``homogeneous'', we mean that the equation admits independent scaling transformations of each independent (and also dependent) variable. Any such algebraic solution, being a manifold of zeros of a homogeneous polynomial in a complex space, determines a compact complex smooth algebraic manifold. It is known for a long time that the complex Monge-Amp\`ere equation has a solution which determines a compact $K3$ surface, which is called $K3$ instanton by physicists \cite{ahs}.  However, it is extraordinarily difficult to find algebraic solutions, and $K3$ in particular, by directly solving $CMA$ because of its inhomogeneity. Here we construct a first example of algebraic solutions that determine Ricci-flat anti-self-dual metrics by solving the general heavenly equation.

In section \ref{sec-lax}, we modify the Lax pair of operators from \cite{fer} so that these operators commute on solutions of (\ref{gener}).

In section \ref{sec-tetrad}, we construct a null tetrad for anti-self-dual Ricci-flat metric governed by
the general heavenly equation.

In section \ref{sec-metric}, we construct basis one-forms and ASD metric determined by solutions of the general heavenly equation.

Though the results of sections \ref{sec-tetrad} and \ref{sec-metric} are straightforward consequences of the general Ashtekar-Jacobson-Smolin
and Mason-Newman theorems \cite{ajs,mn}, we need to specify these results for the general heavenly equation and thus obtain an explicit description
of anti-self-dual gravity in terms of solutions of this equation.

In section \ref{sec-canon}, following the idea from Maciej Dunajski's book \cite{dunaj}, we show how the Lax pair of commuting operators for the general heavenly equation could be derived in a straightforward way.

In section \ref{sec-sol}, we explicitly obtain some examples of algebraic solutions of this equation and also show more general solutions.

In section \ref{sec-real}, we consider such a real cross-section of the general heavenly equation, which specifies
the signature of corresponding real metric to be either Euclidean or neutral (ultrahyperbolic).
By solving reality conditions imposed on the complex solutions obtained in section \ref{sec-sol}, we obtain real solutions as
sets of zeros of homogeneous polynomials in dependent and independent variables.
These solutions determine a real ASD Ricci-flat metric with neutral signature.  More generally, we also have non-polynomial
and functionally invariant solutions with the same property.

In section \ref{sec-sym}, we present all point symmetries of a real version of the general heavenly equation and find that
our solutions, for any values of the parameters that they depend on, are invariant solutions with respect to a certain pair of symmetries of the equation.
Hence, they determine ASD vacuum metrics which have two Killing vectors.

\section{Lax pair}
\setcounter{equation}{0}
 \label{sec-lax}

We start from the Lax pair for the equation (\ref{gener}) presented in \cite{fer}
\begin{eqnarray}
   X_1 &=& u_{34}\partial_1 - u_{13}\partial_4 + \gamma\lambda(u_{34}\partial_1 - u_{14}\partial_3),
\nonumber   \\
   X_2 &=& u_{23}\partial_4 - u_{34}\partial_2 + \beta\lambda(u_{34}\partial_2 - u_{24}\partial_3),
  \label{lax}
\end{eqnarray}
where $\partial_1$ means $\partial/\partial_{z^1}$ and so on. The commutator of these operators
does not vanish on solutions of equation (\ref{gener}):
\begin{eqnarray}
  u_{34}\,[X_1,X_2] &=& \{u_{34}u_{234} - u_{23}u_{344} + \lambda\beta(u_{24}u_{334} - u_{34}u_{234})\}\,X_1
  \nonumber \\
   &+& \{u_{34}u_{134} - u_{13}u_{344} + \lambda\gamma(u_{34}u_{134} - u_{14}u_{334})\}\,X_2 .
\end{eqnarray}
For our purposes, we need a Lax pair that commutes on solutions. It has the form
\begin{equation}
  L_0  = \frac{1}{u_{34}}\,X_1 , \quad  M_0  =  \frac{1}{u_{34}}\,X_2 ,
  \label{commlax}
\end{equation}
so that
\begin{equation}
  [L_0,M_0] = \frac{\lambda}{u_{34}^2}\left\{\left(\Phi_4 - \frac{u_{344}}{u_{34}}\Phi\right)\partial_3
   - \left(\Phi_3 - \frac{u_{343}}{u_{34}}\Phi\right)\partial_4\right\},
  \label{commut}
\end{equation}
where $\Phi=\alpha u_{12}u_{34} + \beta u_{13}u_{24} + \gamma u_{14}u_{23}$ is the left-hand side
of the general heavenly equation, $\Phi_3, \Phi_4$ are partial derivatives of $\Phi$ with respect to $z^3, z^4$
and $[L_0,M_0] = 0$ on solutions.

\section{Null tetrad for anti-self-dual vacuum\\ metric}
\setcounter{equation}{0}
 \label{sec-tetrad}

This section is based on the Ashtekar-Jacobson-Smolin theorem \cite{ajs} (see also Mason and Newman \cite{mn}).
 We will use here the notation and formulation of results from the book of Mason and Woodhouse \cite{MasWood}.
 Let $\Omega$ be a holomorphic function of $z^1,z^2,z^3,z^4$. We denote a null tetrad for the general heavenly equation
 by $W,Z,\tilde{W},\tilde{Z}$ and set
 \begin{equation}\label{LM}
    L = W - \lambda \tilde{Z},\quad M = Z - \lambda \tilde{W}.
 \end{equation}
 Define $\Omega$ by the relations $L_0 = \Omega L$ and $M_0 = \Omega M$ with $\Omega$ yet unknown.
Then
\begin{equation}\label{commLM}
    [L_0,M_0] = [\Omega L,\Omega M] = 0.
\end{equation}
Let $\nu$ be a holomorphic 4-form on a four-dimensional complex manifold with the coordinates $\{z^i\}$,
which should satisfy the conditions
\begin{equation}\label{nucond}
    {\mathscr L}_L(\Omega^{-1}\nu) = {\mathscr L}_M(\Omega^{-1}\nu) = 0,
\end{equation}
where ${\mathscr L}$ denotes Lie derivative.

We note that
\[{\mathscr L}_{L_0}(u_{34}dz^1\wedge dz^2\wedge dz^3\wedge dz^4) = {\mathscr L}_{M_0}(u_{34}dz^1\wedge dz^2\wedge dz^3\wedge dz^4) = 0 ,\]
equivalent to
\begin{equation}\label{LMOm}
    {\mathscr L}_{L}(\Omega u_{34}dz^1\wedge dz^2\wedge dz^3\wedge dz^4) = {\mathscr L}_{M}(\Omega u_{34}dz^1\wedge dz^2\wedge dz^3\wedge dz^4) = 0 .
\end{equation}
Comparing (\ref{nucond}) and (\ref{LMOm}) we deduce that
\begin{equation}\label{nu}
    \nu = \Omega^2u_{34}dz^1\wedge dz^2\wedge dz^3\wedge dz^4
\end{equation}
satisfies condition (\ref{nucond}). According to Proposition 13.4.8 in \cite{MasWood}, if $L$ and $M$
satisfy conditions (\ref{commLM}), (\ref{nucond}) and the normalization condition
\begin{equation}\label{norm}
  24 \nu(W,Z,\tilde{W},\tilde{Z}) = 1,
\end{equation}
then $W,Z,\tilde{W},\tilde{Z}$, defined in (\ref{LM}), is a null tetrad for an ASD vacuum metric. Substituting
$\nu$ from (\ref{nu}) into (\ref{norm}), we obtain $\Omega^2 = \beta\gamma\Delta/u_{34}$,
where $\Delta = u_{13}u_{24} - u_{14}u_{23}$, which determines $\Omega$
\begin{equation}\label{Om}
    \Omega = \sqrt{\frac{\beta\gamma\Delta}{u_{34}}}.
\end{equation}
The 4-form $\nu$ becomes
\begin{equation}\label{nufin}
    \nu = \beta\gamma\Delta dz^1\wedge dz^2\wedge dz^3\wedge dz^4.
\end{equation}
Since $L = \Omega^{-1}L_0$, $M = \Omega^{-1}M_0$ with $L_0$ and $M_0$ defined by (\ref{commlax}),
from the definition of $W,Z,\tilde{W},\tilde{Z}$ in (\ref{LM}) we obtain the explicit form of an ASD tetrad frame
\begin{eqnarray}
 & &  W = \frac{u_{34}\partial_1 - u_{13}\partial_4}{\sqrt{\beta\gamma u_{34}\Delta}}\,,\qquad\quad
  Z = \frac{u_{23}\partial_4 - u_{34}\partial_2}{\sqrt{\beta\gamma u_{34}\Delta}}\,, \nonumber \\
 & &  \tilde{W} = \frac{\sqrt{\beta}(u_{24}\partial_3 - u_{34}\partial_2)}{\sqrt{\gamma u_{34}\Delta}}\,, \quad
\tilde{Z} = \frac{\sqrt{\gamma}(u_{14}\partial_3 - u_{34}\partial_1)}{\sqrt{\beta u_{34}\Delta}}\,,
  \label{WZ}
\end{eqnarray}
which is governed by solutions of the general heavenly equation (\ref{gener}) expressed solely in terms
of the independent parameters $\beta$ and $\gamma$
\begin{equation}\label{geneq}
 (\beta + \gamma)u_{12}u_{34} = \beta u_{13}u_{24} + \gamma u_{14}u_{23}.
\end{equation}

\section{Basis one-forms and ASD metric governed by the general heavenly equation}
\setcounter{equation}{0}
 \label{sec-metric}

The corresponding coframe consists of four 1-forms $\omega^i = \omega^i_j dz^j$ which satisfy
the following normalization conditions
\begin{equation}\label{normal}
    \omega^1(W) = \omega^2(Z) = \omega^3(\tilde{W}) = \omega^4(\tilde{Z}) = 1
\end{equation}
with all other $\omega^i(W), \omega^i(Z), \omega^i(\tilde{W}), \omega^i(\tilde{Z})$ vanishing.
By solving these bi-orthogonality relations, we obtain the following coframe 1-forms
\begin{eqnarray}
  \omega^1 &=& \sqrt{\frac{\beta\gamma}{u_{34}\Delta}}\left\{u_{23}(u_{14}dz^1 + u_{24}dz^2)
  + u_{34}(u_{23}dz^3 + u_{24}dz^4)\right\} \nonumber \\
  \omega^2 &=& \sqrt{\frac{\beta\gamma}{u_{34}\Delta}}\left\{u_{13}(u_{14}dz^1 + u_{24}dz^2)
  + u_{34}(u_{13}dz^3 + u_{14}dz^4)\right\} \nonumber \\
  \omega^3 &=& - \sqrt{\frac{\gamma}{\beta u_{34}\Delta}}\left\{u_{14}(u_{13}dz^1 + u_{23}dz^2)
  + u_{34}(u_{13}dz^3 + u_{14}dz^4)\right\} \nonumber \\
  \omega^4 &=& \sqrt{\frac{\beta}{\gamma u_{34}\Delta}}\left\{u_{24}(u_{13}dz^1 + u_{23}dz^2)
  + u_{34}(u_{23}dz^3 + u_{24}dz^4)\right\}.
  \label{forms}
\end{eqnarray}
On solutions of (\ref{geneq}) the corresponding ASD vacuum metric reads
\begin{eqnarray}
  & & ds^2 = 2(\omega^2\omega^4 - \omega^1\omega^3) = \frac{2(\beta + \gamma)}{\Delta} \times \nonumber \\
  & & \left\{u_{12}\left[u_{13}u_{14}(dz^1)^2 + u_{23}u_{24}(dz^2)^2\right]
  + u_{34}\left[u_{13}u_{23}(dz^3)^2 + u_{14}u_{24}(dz^4)^2\right] \right.\nonumber \\
  & &  \left. \mbox{} + (u_{13}u_{24} + u_{14}u_{23})\left(u_{12}dz^1dz^2 + u_{34}dz^3dz^4\right) \right. \nonumber \\
  & & \left. \mbox{} + (u_{12}u_{34} + u_{14}u_{23})\left(u_{13}dz^1dz^3 + u_{24}dz^2dz^4\right)  \right. \nonumber \\
  & & \left. \mbox{} + (u_{12}u_{34} + u_{13}u_{24})\left(u_{23}dz^2dz^3 + u_{14}dz^1dz^4\right) \right\} .
  \label{metr}
\end{eqnarray}
Using the program EXCALC run by REDUCE, we have computed the Riemann curvature 2-forms and checked vanishing of the Ricci tensor
on solutions of the general heavenly equation in the form (\ref{geneq}), so our metric is indeed Ricci-flat. The expressions
for the Riemann curvature 2-forms are too lengthy to be presented here.

\section{Derivation of the Lax pair}
\setcounter{equation}{0}
 \label{sec-canon}

Here we will show that the Lax pair (defined by (\ref{lax}), (\ref{commlax})) of commuting operators $L_0$ and $M_0$ together with the equation generated by it could be derived in a straightforward way. In doing this, we will follow the idea and notation from the book of Dunajski \cite{dunaj}.

We search for Lax operators linear in the spectral parameter $\lambda$:
\begin{equation}\label{L0L1}
    L_0 = \mathbf{e}_{00'} - \lambda \mathbf{e}_{01'},\quad \rm{and}\quad M_0\equiv L_1 = \mathbf{e}_{10'} - \lambda \mathbf{e}_{11'},
\end{equation}
with the spinor indices. We require $[L_0,M_0] = 0$ for operators (\ref{L0L1}) which by splitting in $\lambda$ yields three equations
\begin{equation}\label{comm_e}
    [\mathbf{e}_{00'},\mathbf{e}_{10'}] = 0,\quad [\mathbf{e}_{00'},\mathbf{e}_{11'}] + [\mathbf{e}_{01'},\mathbf{e}_{10'}] = 0,\quad
    [\mathbf{e}_{01'},\mathbf{e}_{11'}] = 0 .
\end{equation}
We look for $\mathbf{e}_{AA'}$ as linear combinations of partial derivatives operators $\partial_i$ with variable coefficients:
\begin{eqnarray}
   & \mathbf{e}_{00'} = A\partial_1 + B\partial_2 + C\partial_3 + D\partial_4,&\quad \mathbf{e}_{10'} = E\partial_1 + F\partial_2 + G\partial_3 + H\partial_4, \nonumber \\
   & \mathbf{e}_{01'} = a\partial_1 + b\partial_2 + c\partial_3 + d\partial_4,&\quad \mathbf{e}_{11'} = e\partial_1 + f\partial_2 + g\partial_3 + h\partial_4 .
   \label{lincomb}
\end{eqnarray}
Substituting expressions (\ref{lincomb}) into equations (\ref{comm_e}), computing the commutators and separately equating to zero coefficients of the operators $\partial_i$, we obtain 3 groups, each of 4 equations, for the coefficients in (\ref{lincomb}).
The first group of equations reads
\begin{eqnarray}
   & & AE_1 + BE_2 + CE_3 + DE_4 - EA_1 - FA_2 - GA_3 - HA_4 = 0 ,
   \nonumber \\
   & & AF_1 + BF_2 + CF_3 + DF_4 - EB_1 - FB_2 - GB_3 - HB_4 = 0,
   \nonumber \\
   & & AG_1 + BG_2 + CG_3 + DG_4 - EC_1 - FC_2 - GC_3 - HC_4 = 0 ,
   \nonumber \\
   & &  AH_1 + BH_2 + CH_3 + DH_4 - ED_1 - FD_2 - GD_3 - HD_4 = 0 .
   \label{eq4}
\end{eqnarray}
We can always normalize $\mathbf{e}_{00'}$ by choosing $A=1$. To make the first three equations in this group to be identically satisfied, we can choose
$B=C=E=G=0$ and $F=const.$ Later we notice that a convenient choice is $F=-1$. The last equation in (\ref{eq4}) becomes
\begin{equation}\label{eq4a}
    H_1 + D_2 + DH_4 - HD_4 = 0
\end{equation}
whereas the first two vector fields $\mathbf{e}_{AA'}$ reduce to
\begin{equation}\label{e0010}
    \mathbf{e}_{00'} = \partial_1 + D\partial_4,\qquad \mathbf{e}_{10'} = - \partial_2 + H\partial_4.
\end{equation}
With our choice of coefficients $A,B,C,E,F,G$ the second group of four equations in (\ref{comm_e}) becomes
\begin{eqnarray}
   & & e_1 + a_2 + De_4 - Ha_4 = 0,\quad f_1 + b_2 + Df_4 - Hb_4 = 0,\quad
   \label{eq56} \\
   & & g_1 + c_2 + Dg_4 - Hc_4 = 0,
   \label{eq7}  \\
   & & h_1 + d_2 + Dh_4 - hD_4 + dH_4 - Hd_4 \nonumber \\
   & & \mbox{} + aH_1 + bH_2 + cH_3 - eD_1 - fD_2 - gD_3 = 0.
   \label{eq8}
\end{eqnarray}
To satisfy equations (\ref{eq56}) identically, we set constant values to $a = - \gamma$ and $f = - \beta $ and we choose
$b = d = e = h = 0$. Equation (\ref{eq8}) becomes
\begin{equation}\label{eq8a}
    - \gamma H_1 + \beta D_2  + cH_3 - gD_3 =0.
\end{equation}
The third commutator equation in (\ref{comm_e}) for the chosen values of coefficients has only one identically nonvanishing component: the term with $\partial_3$.
Equating it to zero, we obtain
\begin{equation}\label{eq11}
    - \gamma g_1 + \beta c_2 + cg_3 - gc_3 = 0.
\end{equation}
The last two vector fields $\mathbf{e}_{AA'}$ in (\ref{lincomb}) become
\begin{equation}\label{e0111}
    \mathbf{e}_{01'} = - \gamma\partial_1 + c\partial_3,\qquad \mathbf{e}_{11'} = - \beta\partial_2 + g\partial_3.
\end{equation}

Consider now a holomorphic 4-form on a four-dimensional complex manifold with the coordinates $\{z^i\}$
\begin{equation}\label{ro}
    \rho = \Lambda(z) dz^1\wedge dz^2\wedge dz^3\wedge dz^4
\end{equation}
with $\Lambda(z)$ such that $\rho$ should satisfy the conditions
\begin{equation}\label{Lro}
    {\mathscr L}_{L_0}(\rho) = {\mathscr L}_{M_0}(\rho) = 0,
\end{equation}
where ${\mathscr L}$ denotes Lie derivative and $L_0, M_0$ have the form (\ref{L0L1}). With our previous choices of the coefficients in $\mathbf{e}_{AA'}$
equations (\ref{Lro}), being split in $\lambda$, produce the four equations
\begin{eqnarray}
   & & \Lambda_1 + (D\Lambda)_4 = 0,\quad - \Lambda_2 + (H\Lambda)_4 = 0, \nonumber \\
   & & \gamma\Lambda_1 - (c\Lambda)_3 = 0,\quad \beta\Lambda_2 - (g\Lambda)_3 = 0.
   \label{9_10}
\end{eqnarray}
Equations (\ref{9_10}) are satisfied by introducing the potentials $V$ and $W$
\begin{eqnarray}
  & &  \Lambda = V_4,\quad D\Lambda = - V_1,\quad H\Lambda = V_2, \nonumber \\
  & &  \Lambda = W_3,\; c\Lambda = \gamma W_1,\quad g\Lambda = \beta W_2 ,
  \label{VW}
\end{eqnarray}
which implies $V_4 = W_3$. Hence there exists a potential $u$ such that $V = u_3$, $W = u_4$, so that $\Lambda = u_{34}$. Eliminating $\Lambda$ in (\ref{VW}), we obtain
\begin{equation}\label{HD}
    H = \frac{V_2}{V_4} = \frac{u_{23}}{u_{34}},\qquad D = - \frac{V_1}{V_4} = - \frac{u_{13}}{u_{34}},
\end{equation}
\begin{equation}\label{cg}
    c = \gamma\frac{W_1}{W_3} = \gamma\frac{u_{14}}{u_{34}},\qquad g = \beta\frac{W_2}{W_3} = \beta\frac{u_{24}}{u_{34}} ,
\end{equation}
which solve equations (\ref{eq4a}) and (\ref{eq11}). The 4-form (\ref{ro}) becomes 
\[\rho = u_{34}dz^1\wedge dz^2\wedge dz^3\wedge dz^4 = \frac{1}{\Omega^2}\,\nu \]
with the 4-form $\nu$ defined in (\ref{nu}).

The two remaining equations (\ref{eq7}) and (\ref{eq8}) expressed in terms of potential $u$ can be put in the form $\partial_4(\Phi/u_{34}) = 0$ and $\partial_3(\Phi/u_{34}) = 0$ respectively, so that $\Phi = \mu(z^1,z^2)u_{34}$. Here, same as at the end of section \ref{sec-lax}, $\Phi$ denotes the left-hand side of the general heavenly equation (\ref{gener}). By a suitable redefinition of $u$ this becomes $\Phi = 0$, which is exactly the general heavenly equation (\ref{gener}).
Using formulae (\ref{HD}) and (\ref{cg}), we obtain the vector fields $\mathbf{e}_{AA'}$ (\ref{e0010}) and (\ref{e0111}) in the final form
\begin{eqnarray}
  & & \mathbf{e}_{00'} = \partial_1  - \frac{u_{13}}{u_{34}}\partial_4,\qquad\qquad \mathbf{e}_{10'} = - \partial_2 + \frac{u_{23}}{u_{34}}\partial_4 ,
  \nonumber \\
   & & \mathbf{e}_{01'} = - \gamma\left(\partial_1 - \frac{u_{14}}{u_{34}}\partial_3\right),\quad \mathbf{e}_{11'} = - \beta\left(\partial_2 - \frac{u_{24}}{u_{34}}\partial_3\right)
   \label{e_AB}
\end{eqnarray}
and the operators $L_0, M_0$ in (\ref{L0L1}) coincide with the Lax pair (\ref{commlax}) for the general heavenly equation (\ref{gener}).

\section{Algebraic solutions of the general heavenly equation}
\setcounter{equation}{0}
 \label{sec-sol}

 We look for solutions of the general heavenly equation (\ref{gener}), which are algebraic surfaces
 of even order $2m$, $m=1,2,3,\dots$ of the special form
 \begin{eqnarray}
  & & u^{2m} - \left\{a_1(z^1)^{2m} + a_2(z^2)^{2m} + a_3(z^3)^{2m} + a_4(z^4)^{2m} + a_5(z^1)^m(z^2)^m \right.\nonumber \\
  & & \left. \mbox{} + a_6(z^1)^m(z^4)^m + a_7(z^2)^m(z^3)^m) + a_8(z^3)^m(z^4)^m\right. \nonumber \\
   & & \left. \mbox{} + a_9(z^1)^m(z^3)^m + a_{10}(z^2)^m(z^4)^m + C\right\} = 0,
  \label{algsol}
 \end{eqnarray}
 where $C$ is an arbitrary constant and the constant coefficients $a_i$ are determined by equation (\ref{gener}). A complete description
 of this class of solutions is given in the following statement.
 \begin{theorem}
Solution of the form (\ref{algsol}) with $m\not = 1\ \rm{or}\ 2$ to equation (\ref{gener}), which satisfies the condition: $\Delta = u_{13}u_{24} - u_{14}u_{23}\not\equiv 0$
in the denominators of coframe 1-forms (\ref{forms}) and metric (\ref{metr}), exists iff its coefficients are determined
by one of the two sets of relations
 \begin{eqnarray}
  & & a_1 = \frac{a_6(\alpha a_9a_{10}+\beta a_5a_8)+2\gamma a_4a_5a_9}{2\gamma (2a_4a_7-a_8a_{10})}\,,\quad
      a_7 = - \frac{\alpha a_5a_8 + \beta a_9a_{10}}{\gamma a_6}\,,  \nonumber \\
  & & a_2 = \frac{\alpha a_{10}(a_5a_8-a_9a_{10}) + \gamma a_5(2a_4a_7-a_8a_{10})}{2\gamma (2a_4a_9 - a_6a_8)}\,,\nonumber \\
  & & a_3 = \frac{\beta a_8(a_5a_8-a_9a_{10}) - \gamma a_9(2a_4a_7-a_8a_{10})}{2\gamma (a_6a_{10} - 2a_4a_5)}\,,
   \label{sol1}
 \end{eqnarray}
where seven coefficients $a_4,a_5,a_6,a_8,a_9,a_{10},C$ are free parameters, or
\begin{eqnarray}
  & & a_1 = \frac{\beta a_4a_5a_9^2}{\alpha a_5a_8^2 + 2\beta a_4a_7a_9}\,,\quad
      a_2 = \frac{a_5(2\beta^2a_4a_7a_9 - \alpha\gamma a_5a_8^2)}{4\beta^2a_4a_9^2}\,, \nonumber \\
  & & a_3 = \frac{2\beta a_4a_7a_9 - \gamma a_5a_8^2}{4\beta a_4a_5},\,,\quad a_6 = 0,\quad
      a_{10} = - \frac{\alpha a_5a_8}{\beta a_9}
  \label{sol2}
\end{eqnarray}
with six free parameters $a_4,a_5,a_7,a_8,a_9,C$.
\end{theorem}
The proof of this statement has been obtained as an output of an appropriate Reduce program where solutions with
$\Delta\equiv 0$ have been eliminated. It is remarkable that these coefficients do not depend
on the choice of power $m$. In the special cases $m=1$, $m=2$ solutions have more general form which will be studied elsewhere.

Let $z=\{z^1,z^2,z^3,z^4\}$. The proof of the following statement is starightforward.
\begin{theorem}
If $P(z)$ satisfies equation (\ref{gener}) and the differential constraint
\begin{eqnarray}
   & & \alpha(P_1P_2P_{34} + P_3P_4P_{12}) + \beta(P_1P_3P_{24} + P_2P_4P_{13}) \nonumber
 \\
   & & \qquad\qquad\qquad\qquad\quad \mbox{} + \gamma(P_1P_4P_{23} + P_2P_3P_{14}) = 0 ,
      \label{difcon}
\end{eqnarray}
then the function $u(z)$, implicitly determined by the equation $G(u,P) = 0$ where $G$ is an arbitrary smooth  function, is also a solution of (\ref{gener}).
\end{theorem}
Thus, we have obtained a functionally invariant solution. An example of $P(z)$, that satisfies both conditions of this theorem, is given
in curly braces in (\ref{algsol}) with coefficients determined by (\ref{sol1}).
We also note that solutions of the form (\ref{algsol}) are valid more generally for an arbitrary real or complex parameter $m$ but then
they are not algebraic manifolds.

\begin{remark}
The left-hand side of differential constraint (\ref{difcon}) with $P$ replaced by $u$ is the Lagrangian density for the general heavenly equation (\ref{gener}):
\begin{eqnarray}
 L & = & \alpha(u_1u_2u_{34} + u_3u_4u_{12}) + \beta(u_1u_3u_{24} + u_2u_4u_{13}) \nonumber
 \\
   & & \qquad\qquad\qquad\qquad\quad \mbox{} + \gamma(u_1u_4u_{23} + u_2u_3u_{14}) ,
      \label{lagrange}
\end{eqnarray}
i.e. equation (\ref{gener}) is the Euler-Lagrange equation for Lagrangian (\ref{lagrange}). Thus, the differential constraint (\ref{difcon})
selects those solutions of (\ref{gener}) for which $L = 0$.
\end{remark}

The following remark is relevant to a characterization of the joint solution space of the two equations involved in proposition\ \thetheorem.
To be specific, we will use equations (\ref{gener}) and (\ref{difcon}) algebraically solved with respect to $u_{12}$ and $u_{13}$.
\begin{remark}
The system of two equations (\ref{gener}) and (\ref{difcon}) (with $P$ replaced by $u$) has two third-order integrability conditions
\begin{eqnarray}
 & &   A (u_2u_{34} - u_3u_{24}) + B (u_4u_{23} - u_3u_{24}) = 0, \nonumber \\
 & &   D (u_2u_{34} - u_3u_{24}) + E (u_4u_{23} - u_3u_{24}) = 0,
 \label{ord3rd}
\end{eqnarray}
where
\begin{eqnarray*}
 \hspace*{-5mm} & & A = u_4(u_4u_{223} - u_2u_{234}) - u_3(u_4u_{224} - u_2u_{244}) + u_2(u_{24}u_{34} - u_{23}u_{44}), \\
 \hspace*{-5mm} & & B = u_2(u_2u_{344} - u_4u_{234}) - u_3(u_2u_{244} - u_4u_{224}) + u_4(u_{23}u_{24} - u_{22}u_{34}), \\
 \hspace*{-5mm} & & D = 2u_3u_4u_{234} - u_3^2u_{244} - u_4^2u_{233} + u_{23}(u_{3}u_{44} - u_{4}u_{34}) - u_{24}(u_{3}u_{34} - u_{4}u_{33}) \\
 \hspace*{-5mm} & & E = u_3(u_3u_{244} - u_4u_{234}) - u_2(u_3u_{344} - u_4u_{334}) + u_4(u_{23}u_{34} - u_{24}u_{33}) .
\end{eqnarray*}
Applying the algorithms of \cite{kur,yan}, one can check that the system of four equations (\ref{gener}), (\ref{difcon}) and (\ref{ord3rd}) is involutive (generates no further integrability conditions) and that
the main part of arbitrariness of its  solution manifold is determined by three arbitrary functions of two variables,
whereas the general solution manifold of equation (\ref{gener}) without differential constraints depends on two arbitrary functions of three variables.
\end{remark}
In the generic case, when $(u_2u_{34} - u_3u_{24}, u_4u_{23} - u_3u_{24})\not = (0, 0)$, we obtain the constraint
\begin{equation}\label{det}
    \det\left[
    \begin{array}{cc}
    A & B \\
    D & E
    \end{array}
    \right] = 0.
\end{equation}
Two equations (\ref{ord3rd}) become linearly dependent, so we still have two third-order integrability conditions: one equation in (\ref{ord3rd})
and (\ref{det}). No further integrability conditions are generated for the system (\ref{gener}), (\ref{difcon}), (\ref{ord3rd}) and (\ref{det}).

\section{Real cross-section of the general heavenly equation and its real algebraic solutions}
\setcounter{equation}{0}
 \label{sec-real}

For applications to self-dual gravity we need real cross-sections of the general heavenly equation and its solutions.
We specify the real cross-section by the requirement that the corresponding real metric should have a certain signature.-
Then we have to make the following identifications: $z^2 = \bar{z}^1$ and $z^4 = \bar{z}^3$ ($z^2 = - \bar{z}^1$ would also do)
and then replace everywhere index 3 by 2. Here the bar means complex conjugation.

The real general heavenly equation takes the form
\begin{equation}
    \alpha u_{1\bar 1}u_{2\bar 2} + \beta u_{12}u_{\bar 1\bar 2} + \gamma u_{1\bar 2}u_{2\bar 1} = 0,
    \label{realheav}
\end{equation}
while $\Delta = u_{12}u_{\bar 1\bar 2} - u_{1\bar 2}u_{2\bar 1}$. In the following we assume that $\gamma\not=0$.

The signature of the metric depends on the sign of $\beta/\gamma$. If $\beta/\gamma > 0$, we set
$\beta = \gamma\delta^2$ with $\delta > 0$. Then basis 1-forms (\ref{forms}) become
\begin{eqnarray}
   & & \omega^1 = \frac{|\gamma|\delta}{\sqrt{u_{2\bar 2}\Delta}}\,\bar{l}_1,\qquad
   \omega^2 = \frac{|\gamma|\delta}{\sqrt{u_{2\bar 2}\Delta}}\,l_2,\nonumber \\
   & & \omega^3 = - \frac{1}{\delta\sqrt{u_{2\bar 2}\Delta}}\,l_1,\qquad
   \omega^4 = \frac{\delta}{\sqrt{u_{2\bar 2}\Delta}}\,\bar{l}_2,
   \label{om_euclid}
\end{eqnarray}
where 1-forms $l_1$ and $l_2$ are defined as
\begin{eqnarray}
  l_1 &=& u_{1\bar 2}(u_{12}dz^1 + u_{\bar 12}d\bar{z}^1) + u_{2\bar 2}(u_{12}dz^2 + u_{1\bar 2}d\bar{z}^2), \\
  l_2 &=& u_{12}(u_{1\bar 2}dz^1 + u_{\bar 1\bar 2}d\bar{z}^1) + u_{2\bar 2}(u_{12}dz^2 + u_{1\bar 2}d\bar{z}^2).
  \label{l}
\end{eqnarray}
The metric becomes
\begin{equation}
  ds^2 = 2(\omega^2\omega^4 - \omega^1\omega^3) = \frac{2|\gamma|}{|u_{2\bar 2}\Delta|}
  \left\{|l_1|^2 + \delta^2|l_2|^2\right\},
  \label{euclidmetr}
\end{equation}
which has obviously Euclidean signature. It is determined by solutions of the real version (\ref{realheav})
of the general heavenly equation in the form
\begin{equation}
   (\delta^2 + 1)u_{1\bar 1}u_{2\bar 2} =  \delta^2 u_{12}u_{\bar 1\bar 2} + u_{1\bar 2}u_{2\bar 1}.
   \label{euclideq}
\end{equation}

If $\beta/\gamma < 0$, we set $\beta = - \gamma\delta^2$ and then the real cross-section of the
general heavenly equation (\ref{realheav}) becomes
\begin{equation}
   (\delta^2 - 1)u_{1\bar 1}u_{2\bar 2} = \delta^2 u_{12}u_{\bar 1\bar 2} - u_{1\bar 2}u_{2\bar 1}.
   \label{neutraleq}
\end{equation}
Basis 1-forms (\ref{forms}) become
\begin{eqnarray}
      & & \omega^1 = \frac{i|\gamma|\delta}{\sqrt{u_{2\bar 2}\Delta}}\,\bar{l}_1,\qquad
   \omega^2 = \frac{i|\gamma|\delta}{\sqrt{u_{2\bar 2}\Delta}}\,l_2,\nonumber \\
   & & \omega^3 = - \frac{1}{i\delta\sqrt{u_{2\bar 2}\Delta}}\,l_1,\qquad
   \omega^4 = \frac{i\delta}{\sqrt{u_{2\bar 2}\Delta}}\,\bar{l}_2,
   \label{realomeg}
\end{eqnarray}

The metric (\ref{metr}) takes the form
\begin{equation}
  ds^2 = 2(\omega^2\omega^4 - \omega^1\omega^3) = \frac{2|\gamma|}{|u_{2\bar 2}\Delta|}
  \left\{|l_1|^2 - \delta^2|l_2|^2\right\},
  \label{realmetr}
\end{equation}
which obviously has neutral signature. This metric is determined by solutions of equation (\ref{neutraleq}).

So far, we were able to obtain a real solution only in the case of neutral signature. This solution
of equation (\ref{neutraleq}) is obtained by solving reality conditions imposed on
the complex algebraic solution (\ref{algsol}) with coefficients~(\ref{sol1}). A relatively simple particular solution has the following form:
\begin{eqnarray}
  & & u^{2m} - \left\{\frac{A}{2} \frac{(F-\delta R)}{(\delta F-R)}\left[\exp{(2i\theta)}(z^1)^{2m}+\exp{(-2i\theta)}({\bar z}^1)^{2m}\right] \right.
  \nonumber \\
  & &  \left. \mbox{} + \frac{1}{2A}\left[\delta(AB+F^2)-FR\right] \left[(z^2)^{2m}+({\bar z}^2)^{2m}\right] - A(z^1)^m({\bar z}^1)^m\right.
  \nonumber \\
  & &  \left. \mbox{} + R\left[\exp{(i\theta)}(z^1)^m({\bar z}^2)^m+\exp{(-i\theta)}(z^2)^m({\bar z}^1)^m\right]  + B(z^2)^m({\bar z}^2)^m \right.
  \nonumber \\
  & &  \left. \mbox{} + F\left[\exp{(i\theta)}(z^1)^m(z^2)^m+\exp{(-i\theta)}({\bar z}^1)^m({\bar z}^2)^m\right] + C\right\} = 0,
  \label{realsol}
\end{eqnarray}
where $R = \sqrt{(\delta^2-1)AB + \delta^2F^2}$ and the constants $m,A,B,F,\theta,C$ are free real parameters (obviously, from now on we use $A,B,C,E,F$ in a different sense than in the preceding sections). Parameter $\theta$ is inessential,
since it can be scaled out by using scaling symmetries of the general heavenly equation (\ref{neutraleq}), so that we can set $\theta=0$
in (\ref{realsol}) and our solution depends only on five essential parameters. For $m=1,2,3,\dots$ this solution determines an algebraic hypersurface
in the real five-dimensional space.

The real cross-section of the second complex solution (\ref{algsol}), (\ref{sol2}) can be transformed to a particular case of solution
(\ref{realsol}) by using scaling symmetries of the equation (\ref{neutraleq}) and therefore does not yield an essentially new real solution.

A general solution of reality conditions for (\ref{algsol}) with coefficients~(\ref{sol1}) reads
\begin{eqnarray}
  & & u^{2m} - \left\{a_1(z^1)^{2m} + \bar{a}_1(\bar{z}^1)^{2m} + E\left[(z^2)^{2m} + (\bar{z}^2)^{2m}\right] - A(z^1)^m(\bar{z}^1)^m \right.\nonumber \\
  & & \left. \mbox{} + R\left[(z^1)^m(\bar{z}^2)^m + (z^2)^m(\bar{z}^1)^m)\right] + B(z^2)^m(\bar{z}^2)^m\right. \nonumber \\
   & & \left. \mbox{} + F\left[e^{i\phi}(z^1)^m(z^2)^m + e^{-i\phi}(\bar{z}^1)^m(\bar{z}^2)^m\right] + C\right\} = 0,
  \label{genersol}
 \end{eqnarray}
where all inessential parameters are scaled out by scaling symmetries of the general heavenly equation (\ref{neutraleq}), the coefficients satisfy the relations
\begin{equation}\label{a_1}
    a_1 = \frac{2AEFe^{i\phi} - R(R^2 + AB - F^2)}{2(BFe^{-i\phi} - 2ER)}\,,
\end{equation}
\begin{equation}\label{theta}
    \cos{\phi} = \frac{B(R^2 + AB + F^2) - 4AE^2}{4EFR},
\end{equation}
$R$ is defined above and $\bar{a}_1$ is complex conjugate to (\ref{a_1}). This solution at $m=1,2,3,\dots$ and fixed $\delta$ depends on five real parameters, four of which,
$A, B, E, F$, should satisfy the relation
\begin{equation}
  |B(R^2 + F^2 + AB) - 4AE^2|\leqslant 4|EF|R
\label{inequ}
\end{equation}
to ensure $|\cos{\phi}|\leqslant 1$. It is not difficult to check that inequality (\ref{inequ}) can indeed be satisfied for the considered
case of neutral signature, when $\beta/\gamma = - \delta^2 < 0$, while for $\beta/\gamma = \delta^2 > 0$ in the case of Euclidean signature
it cannot be satisfied for real solutions.

  The particular solution (\ref{realsol}) with $\theta=0$ is obtained from (\ref{genersol}) at $\phi=0$, which implies the relation $E = \frac{\displaystyle\pm\delta|AB + F^2| - FR}{\displaystyle 2A}$.\vspace{1mm}

Using our solutions (\ref{realsol}), (\ref{genersol}) in metric (\ref{realmetr}) together with definitions (\ref{l}) of 1-forms $l_1,l_2$,
we obtain real ASD Ricci-flat metric with neutral signature, which depends on five real parameters. We have checked that $\Delta\not\equiv 0$
as far as $A\cdot (AB+F^2)\cdot (\delta^2-1)\not= 0$ (for $m=1$ also $C\ne 0$) while
$u_{2\bar 2}\not=0$ obviously implies $B\ne 0$. Under these conditions, metric (\ref{realmetr}) has no identically vanishing denominators.

\section{Symmetries of general heavenly equation}
\setcounter{equation}{0}
 \label{sec-sym}

We have determined all point symmetries of the general heavenly equation. For its real version (\ref{neutraleq}), which governs the ASD metrics
with the neutral signature, symmetry generators have the form
\begin{eqnarray}
   & & X_1 = a(z^1)\partial_1,\;  \bar{X}_1 = \bar{a}(\bar{z}^1)\partial_{\bar 1},\quad X_2 = b(z^2)\partial_2,\;  \bar{X}_2 = \bar{b}(\bar{z}^2)\partial_{\bar 2}, \; X_3 = u\partial_u, \nonumber \\
   & &   X_4 = c(z^1)\partial_u,\;  \bar{X}_4 = \bar{c}(\bar{z}^1)\partial_u,\quad X_5 = d(z^2)\partial_u,\;  \bar{X}_5 = \bar{d}(\bar{z}^2)\partial_u , \label{symgen}
\end{eqnarray}
where $a,b,c,d$ and their complex conjugates are arbitrary functions of one variable. Finite symmetry transformations generated by $X_3, X_4, X_5$ and
$\bar{X}_4, \bar{X}_5$ in (\ref{symgen})
incorporate scaling $\tilde{u} = \lambda u$ and translations $\tilde{u} = u + \varepsilon c(z^i)$ for $i=1,2$, together with their complex conjugates.
Symmetry transformations generated by $X_1, X_2$ have the form $\hat{a}(\tilde{z}^i) = \hat{a}(z^i) + \varepsilon$, where we have introduced the notation $\hat{a}(z) = \int\frac{\displaystyle dz}{\displaystyle a(z)}$, plus complex conjugate equations.  In particular, if either $a(z) = 1$ or $a(z) = z$
and similarly for their complex conjugates, we obtain translations and scaling in each variable $z^i, \bar{z}^i$ respectively.
We note that our solutions (\ref{realsol}) and (\ref{genersol}) with $C\ne 0$ are noninvariant under these particular cases of symmetry transformations. However, consider the invariance condition of our solution (\ref{genersol}) under the symmetry generator $X = X_1 + X_2 + \bar{X}_1 + \bar{X}_2$ in (\ref{symgen}), with the choice $a(z^1) = c_1/(z^1)^{m-1}, b(z^2) = c_2/(z^2)^{m-1}$ and their complex conjugates:
\begin{eqnarray}
   & & 2a_1c_1 - A\bar{c}_1 + Fe^{i\phi}c_2 + R\bar{c}_2 = 0, \nonumber \\
   & & - Ac_1 + 2\bar{a}_1\bar{c}_1 + Rc_2 + Fe^{-i\phi}\bar{c}_2 = 0, \nonumber \\
   & &  Fe^{i\phi}c_1 + R\bar{c}_1 + 2Ec_2 + B\bar{c}_2 = 0, \nonumber \\
   & &  Rc_1 + Fe^{-i\phi}\bar{c}_1 + Bc_2 + 2E\bar{c}_2 = 0 .
   \label{invar}
\end{eqnarray}
This system of linear algebraic equations admits nonzero solution for the coefficients $c_i, \bar{c}_i$ since the determinant of this system turns out to be zero. Moreover, the rank of this system equals 2.
 This proves the existence of two independent symmetries of our equation, such that our real solution of the general form is invariant with respect to these symmetries for any choice of parameters in the solution and hence it is an invariant solution.
  Therefore, the corresponding metric (\ref{realmetr}) will have two Killing vectors.

\section{Conclusion}

Applying the general Ashtekar-Jacobson-Smolin-Mason-Newman theorem, we have obtained an explicit description of anti-self-dual gravity in terms of solutions to the general heavenly equation, introduced by Doubrov and Ferapontov \cite{fer}.
 We have derived the ASD Ricci-flat vacuum metric, determined by solutions of the general heavenly equation, together
with the corresponding null tetrad and basis 1-forms. Following the ideas of Dunajski's book \cite{dunaj}, we have also been able to derive straightforwardly a Lax pair of commuting operators. Unlike other heavenly equations that describe anti-self-dual gravity, the general heavenly equation
is homogeneous, i.e. admit independent scaling transformations of each independent (and also dependent) variable. This property allows us to obtain
algebraic solutions to this equation in the form of homogeneous polynomials in independent and dependent variables, which can be modified by
adding an arbitrary constant $C$. In this respect, the homogeneity
 seems to be most important property of the considered equation, so that we would suggest to call it \textit{homogeneous heavenly equation}.
Though for $C\not=0$, these solutions do not admit scaling or any other obvious symmetries of the equation, we have proved that
they are invariant solutions with respect to certain two symmetries of this equation, so that the corresponding metric will have two Killing vectors.
 The work on noninvariant algebraic solutions to heavenly equation, such that their real form will determine metrics with Euclidean signature, is currently in progress. Such a solution may be relevant to the search of the famous $K3$ instanton \cite{ahs}.

\end{document}